\documentclass[aps,prb,twocolumn,showpacs]{revtex4-1}

\usepackage{graphicx}
\usepackage{calc}
\usepackage{bm}
\usepackage{color}
\usepackage{textcomp}

\begin{document}

\title{Magnetically stable zero-bias anomaly in Andreev contact to the magnetic Weyl semimetal 
Co$_3$Sn$_2$S$_2$}

\author{O.O.~Shvetsov}
\author{Yu.S.~Barash}
\author{S.V.~Egorov}
\author{A.V.~Timonina}
\author{N.N.~Kolesnikov}
\author{E.V.~Deviatov}
\affiliation{Institute of Solid State Physics of the Russian Academy of Sciences, Chernogolovka, Moscow District, 2 Academician Ossipyan Str., 142432 Russia}

\date{\today}

\begin{abstract}
Being encouraged by the interplay between topology, superconductivity and magnetism, we experimentally 
investigate charge transport through the interface between the Nb superconductor and the time-reversal 
symmetry breaking Weyl semimetal  Co$_3$Sn$_2$S$_2$. In addition to the proximity induced superconducting gap, we 
observe several subgap features, among which the most interesting is the prominent subgap zero-bias anomaly, absolutely stable against external magnetic fields up to the critical field of Nb.  As the promising 
scenario for the zero-bias anomaly to appear in transport characteristics, we consider the proximity induced zero-energy 
Andreev bound states interfaced with the half-metallic Co$_3$Sn$_2$S$_2$ and influenced by the strong 
spin-orbit coupling and large Zeeman splitting.
\end{abstract}


\maketitle

\section{Introduction}

A magnetic Weyl semimetal (WSM) with broken time-reversal symmetry can be interpreted as a 3D heterostructure 
consisting of the layers of Chern insulators~\cite{burkov1}. For WSM, the coupling between the layers closes 
the bulk band gap at isolated points of the Brillouin zone. These band touching points with linear dispersion, 
also called Weyl nodes, are topological objects~\cite{armitage,mag1}. Their topological protection is 
guaranteed by a non-zero Chern number defined on a sphere in momentum space enclosing a given Weyl node. 

In transport studies, an anomalous Hall effect (AHE) is the hallmark of a WSM phase~\cite{burkov2}. 
AHE originates from the topologically protected chiral surface states residing in the Fermi 
arcs, which connect projections of Weyl nodes on the surface Brillouin zone and inherit the chiral property of
Chern insulator edge states~\cite{armitage,mag1}. The Fermi-arc states can play an important role in forming 
the transport properties of WSM not only when the bulk contribution is strongly suppressed or absent for a 
special reason, as in AHE. The anomalous contribution to a number of transport coefficients from the Fermi 
arcs can be of the same order as from the bulk states even in large systems~\cite{brouwer}.

Physics becomes even more exciting if one considers a magnetic WSM in proximity to an $s$-wave spin-singlet
superconductor. The unusual band structure of WSM and its nontrivial topological properties modify Andreev
reflection processes and result in new proximity-induced effects as compared to what was known until 
recently~\cite{andreev,tinkham,tanaka,flensberg,alicea,beenakker,sato,odd_freq}. A variety of possible 
superconducting phases with and without Weyl points, that can possess distinct topological features depending
on the Zeeman splitting strength and the magnitude of the proximity effect, have been theoretically identified
within the BCS-like model in magnetic WSM films \cite{meng}. In this case the superconductivity splits each 
electronic state into a particle-hole symmetric and antisymmetric states. Correspondingly, the system does not
generally develop a superconducting gap, but splits each Weyl node of a given chirality into two separated 
Bogoliubov-Weyl nodes with opposite particle-hole symmetry. The particle-hole symmetric and antisymmetric 
subspaces are decoupled and the Majorana bound states can eventually emerge at the WSM surface. On 
the other hand, conventional Andreev reflection should be suppressed for the bulk excitations near a 
certain Weyl node due to chirality blockade~\cite{chblock}: a change in chirality is required in order 
to preserve spin-singlet pairing with its transfer of zero-spin and zero-momentum. When the proximity-induced
superconductivity in the bulk is substantially suppressed, the superconductivity in the Fermi arcs has been 
predicted to show up and, in the simplest cases, found to emerge as a pure triplet state or a singlet-triplet 
mixture~\cite{arcsc}. The appearance of chiral zero-energy Majorana bound states has been identified for the 
triplet pairing. Thus, topologically protected surface states of WSM form a potential platform for realization
of chiral Majorana bound states~\cite{armitage}. 

\begin{figure}
    \includegraphics[width=\columnwidth]{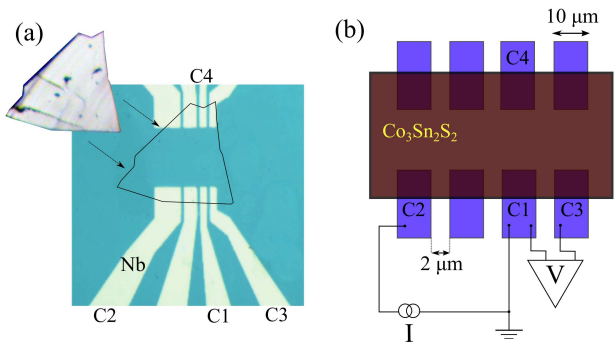}
    \caption{(Color online) (a) The top-view image of the sample. A flat Co$_3$Sn$_2$S$_2$ single crystal 
(about 100~$\mu$m size and 1$\mu$m thick) is transferred on top of the Nb leads, preliminary defined on a 
SiO$_2$ substrate. Planar junctions are formed with $\approx$ 10 $\mu$m overlap. (b) The sketch of a sample 
with the circuit diagram. The distance between the leads is 2~$\mu$m.  Charge transport is investigated with a 
standard three-point technique: the studied contact (C1) is grounded and the two other contacts (C2 and C3) 
are used for applying current and measuring potential.}
    \label{nbcosns_sample}
\end{figure}

There are only a few candidates of time-reversal symmetry breaking WSMs~\cite{mag1,mag2,mag3,mag4}. Recently, 
the giant AHE has been reported~\cite{kagome,kagome1} for the kagome-lattice half-metallic ferromagnet 
Co$_3$Sn$_2$S$_2$ as an anomalous Hall conductance in zero magnetic field. The existence of Fermi arc surface 
states in the material was confirmed by angle-resolved photoemission spectroscopy~\cite{kagome1,kagome2} and 
scanning tunneling microscopy~\cite{kagome_arcs}. The magnetic moments of cobalt become
ferromagnetically ordered out of kagome-lattice Co$_3$Sn planes below 175K. The Co planes are interleaved with
buffer planes of triangularly ordered tin and sulfur. Though the bulk-edge correspondence generally 
relates the topological invariants in the bulk to the topologically protected surface arcs, specific 
connectivities among the Weyl nodes and the corresponding STM/STS spectra turn out to be not generally 
predetermined and can be associated with the particular Sn-, S-, or Co- surface 
terminations~\cite{kagome_arcs,terraces}. 

Here, we experimentally investigate charge transport through the interface between a Nb superconductor and a 
magnetic WSM Co$_3$Sn$_2$S$_2$. Aside from Andreev reflection, we observe several additional features, among 
which the most impressive is the prominent zero-bias anomaly (ZBA), fairly stable against external magnetic 
fields up to the critical field of Nb. We discuss possible scenarios that could result in magnetically stable
ZBA under the conditions studied, taking into account the topological nature of WSM and its interfaces as well
as the presence of the spin-orbit coupling and large Zeeman splitting, which are intrinsic for the 
half-metallic WSM.

\section{Samples and technique}

\begin{figure}
    \includegraphics[width=\columnwidth]{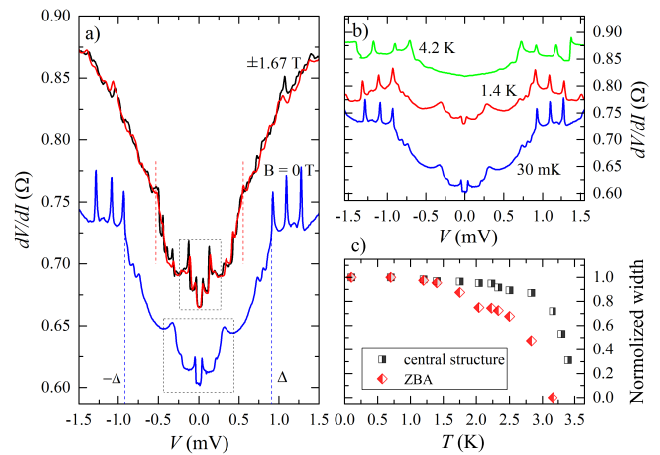}
    \caption{(Color online) $dV/dI(V)$ characteristics in different conditions for one of the 
Nb-Co$_3$Sn$_2$S$_2$ junctions. (a) $dV/dI(V)$ curves are measured at 30~mK. External magnetic field $B$ is 
zero for the blue curve, and equal to $\pm$1.67~T for the black and red curves, respectively. The 
superconducting gap, marked by the blue (zero field) and red ($\pm$1.67~T) dashed lines, is suppressed in 
magnetic fields. The subgap structure is marked by the black dashed rectangle, it is diminishing 
proportionally to the superconducting gap in magnetic fields. ZBA, on the contrary, is robust against 
magnetic fields. Magnetic field is normal to the sample's plane. (b) $dV/dI(V)$ characteristics of the same 
junction at different temperatures. ZBA is suppressed by temperature, while the proximity induced 
superconducting gap is still observable. The curves are shifted for clarity both in (a) and (b). (c) 
Normalized width of the subgap structure (the black rectangles) and ZBA (the red diamonds) versus 
temperature.}
    \label{2n_1}
\end{figure}

Co$_3$Sn$_2$S$_2$ single crystals were grown by the gradient freezing method. An initial load of high-purity 
elements taken in stoichiometric ratio was slowly heated up to 920$^\circ$~C in the horizontally positioned 
evacuated silica ampule, held for 20 h and then cooled with the furnace to ambient temperature at the rate
of 20 deg/h. The obtained ingot was cleaved in the middle part. The Laue patterns confirm the hexagonal 
structure with $(0001)$ as a cleavage plane.  Electron probe microanalysis of cleaved surfaces and x-ray 
diffractometry of powdered samples confirmed the stoichiometric composition of the crystal.

The kagome-lattice ferromagnet Co$_3$Sn$_2$S$_2$ has the (0001) cleavage plane, but bonds between the layers 
in the crystal are quite strong. Thin flakes of Co$_3$Sn$_2$S$_2$ can not be easily mechanically exfoliated, 
e.g., with a scotch-tape technique. Thus, only rather thick flakes (about 100~$\mu$m size and 1$\mu$m thick) 
can be exfoliated by rough mechanical cleavage. We selected the flattest flakes with a clean surface, which 
was verified by an optical microscope.

\begin{figure*}
    \center{\includegraphics[width=300pt]{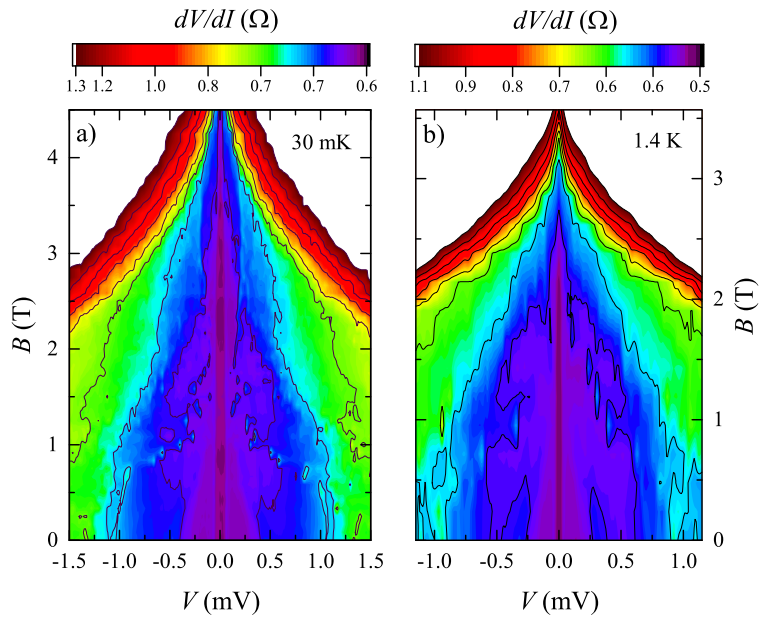}}
    \caption{(Color online) Evolution of $dV/dI(V)$ characteristic of the Nb-Co$_3$Sn$_2$S$_2$ junction in 
magnetic fields (a) at $T$ = 30~mK and (b) at $T$ = 1.4~K. At both temperatures,  ZBA (the violet stripe 
around zero bias) is stable against magnetic fields. Suppression of  ZBA  starts only close to the critical 
field of the superconducting gap.}
    \label{2n_color}
\end{figure*}

Then, a small Co$_3$Sn$_2$S$_2$ flake is transferred to the Nb leads pattern, see Fig.~\ref{nbcosns_sample} 
(a). The leads pattern is defined on an insulating SiO$_2$ substrate by a lift-off technique after magnetron 
sputtering of 150~nm Nb. The leads are separated by 2~$\mu$m intervals. Further, a Co$_3$Sn$_2$S$_2$ flake is 
pressed to the leads slightly with another oxidized silicon substrate. A weak pressure is applied with a 
special metallic frame, which keeps the substrates strictly parallel. This procedure provides transparent 
contacts, stable in different cooling cycles, without mechanical polishing or chemical 
treatment~\cite{nbwte,cdas,magnon}.

It has been shown previously, that our samples demonstrate a giant AHE~\cite{magnon}, which is a hallmark of 
Co$_3$Sn$_2$S$_2$ Weyl semimetal~\cite{kagome,kagome1}. The bulk of the samples is fully spin-polarized~\cite{magnon} above 
0.5~T.

A standard three-point technique is used to study electron transport across a single Nb-Co$_3$Sn$_2$S$_2$ 
junction, see Fig.~\ref{nbcosns_sample}(b): one Nb contact (C1) is grounded and two other contacts are used 
for applying current (C2) and measuring potential (C3). To obtain $dV/dI(V)$ characteristics, dc current is 
additionally modulated by a low (below 5~$\mu$A, $f$ $\approx$ 1~kHz) ac component. We measure both dc ($V$) 
and ac ($\sim dV/dI$) components of the potential with a dc voltmeter and a lock-in, respectively, after a 
broad-band preamplifier. We checked that the lock-in signal is independent of the modulation frequency.
The obtained $dV/dI(V)$ characteristics are verified to be independent of mutual positions of current/voltage 
probes (e.g., changing C2 to C4 or C3 to C4 in Fig.~\ref{nbcosns_sample}(b)), so they reflect transport 
parameters of the Nb-Co$_3$Sn$_2$S$_2$ interface without admixtures of the sample's bulk resistance. We also 
checked that $dV/dI(V)$ characteristics are well reproducible in different cooling cycles. The measurements 
are performed in a dilution refrigerator (30~mK - 1.2~K) and in a usual He$^4$ cryostat (1.4~K - 4.2~K).

\section{Experimental results}

The blue curve in Fig.~\ref{2n_1} demonstrates $dV/dI(V)$ characteristic of the Nb-Co$_3$Sn$_2$S$_2$ junction 
at the lowest temperature 30~mK and zero external magnetic field. The overall behavior of the curve is 
consistent with the known one for a single Andreev contact to a ferromagnet~\cite{sfpc}: niobium 
superconducting gap is observed as a $dV/dI$ drop and denoted by the blue dashed lines. The gap $\Delta$ = 
0.9~meV is the Nb gap suppressed by the ferromagnetism of the Co$_3$Sn$_2$S$_2$ flake in a vicinity of the 
interface by 40\%, as compared to the Nb gap in the bulk $\Delta_{Nb}$ = 1.5~meV. Such a suppression could 
indicate to a large spin-mixing angle acquired by the quasiparticles in the reflection and/or transmission 
interface processes \cite{rainer1988,eschrig2017}. As the subgap resistance is lower than the normal one, one 
may assume that we have a transparent Nb-Co$_3$Sn$_2$S$_2$ interface. Although the interface quality allows us
to observe the Andreev physics in the Nb-Co$_3$Sn$_2$S$_2$ junctions, it prevents the proximity-induced 
Josephson current~\cite{cosnsjc} to flow across the WSM Co$_3$Sn$_2$S$_2$ connecting 2~$\mu$m spaced 
superconducting Nb contacts, irrespective of the presence or absence of the applied external magnetic field. 

In addition to the Nb superconducting gap, the blue curve at zero magnetic field and 30~mK in 
Fig.~\ref{2n_1}(a) demonstrates several features both inside and outside the gap. There are sharp periodic 
peaks outside the Nb gap, which are likely to be geometrical resonances, known for Andreev 
contacts~\cite{bite,tomasch,mcmillan}. Also, there is a wide ($\pm$0.45~meV width) central structure, which is
denoted by the dashed rectangle in Fig.~\ref{2n_1}(a) and can be understood~\cite{indgap1,indgap2,indgap3} as 
the proximity induced soft~\cite{takei2013} superconducting gap $\Delta_S$ at the surface of Co$_3$Sn$_2$S$_2$.  The gap $\Delta_S$ has approximately the BCS-like temperature dependence~\cite{tinkham} 
and nearly satisfies the BCS relation $\Delta_S$ $\approx$ 1.76$k_B T_{c_{S}}$, where $T_{c_{S}}$ = 3.5~K in 
Fig.~\ref{2n_1}(c) and $\Delta_S$ = 0.45~meV in Fig.~\ref{2n_1}(a).

In the center of the subgap structure ZBA appears as a narrow ($\pm$0.06~meV width) $dV/dI$ drop.
Our most notable result is that  ZBA exhibits absolute robustness to external magnetic fields, which is 
demonstrated in Fig.~\ref{2n_1}(a), see the black and red curves at $\pm$1.67~T. The resonances outside the 
gap disappear under the magnetic field, while the subgap structure region, denoted by the black dashed 
rectangle in Fig.~\ref{2n_1}(a), diminishes proportionally to the superconducting gap in the field. This 
behavior is independent on the field's sign; compare the black (+1.67~T) and red (-1.67~T) curves in 
Fig.~\ref{2n_1}(a).

In contrast to the magnetic field, an increase in temperature has a dramatic influence on  ZBA: it shrinks
and disappears completely at higher temperatures, while the superconducting gap is still observable, see 
Fig.~\ref{2n_1}(b). Fig.~\ref{2n_1}(c) demonstrates the temperature dependence of the central structure region
normalized width (the black rectangles) and ZBA normalized width (red diamonds). ZBA completely disappears at 
$T$ = 3.15~K, while the wide subgap structure vanishes at about 3.5~K.

To investigate stability of  ZBA against magnetic fields more precisely, we conducted two detailed 
measurements of $dV/dI(V)$ versus $B$ at two temperatures - the minimal 30~mK and 1.4~K. The results are 
demonstrated as colormaps in Fig.~\ref{2n_color}(a,b). At both temperatures, the width and depth of  ZBA 
are robust against magnetic fields. Suppression of  ZBA starts only close to the critical field of the 
superconducting gap: after 3.3~T at 30~mK (Fig.~\ref{2n_color}(a)) and after 2.2~T at 1.4~K 
(Fig.~\ref{2n_color}(b)). 

\begin{figure}
    \includegraphics[width=\columnwidth]{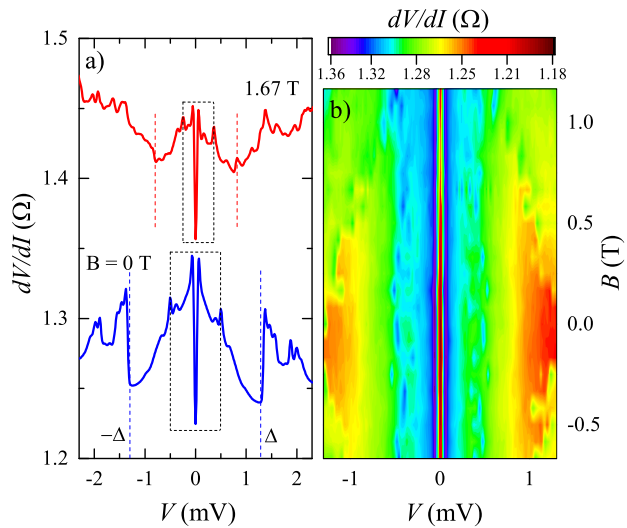}
    \caption{(Color online) $dV/dI(V)$ characteristic for another Nb-Co$_3$Sn$_2$S$_2$ junctions. 
(a) $dV/dI(V)$ curves are measured at 30~mK for zero external magnetic field (blue) and for $B$ = 1.67~T.  
The width of ZBA stays unaltered in high magnetic fields, while it height is slightly diminished. (b) 
Evolution of $dV/dI(V)$ characteristic in magnetic fields. The width of ZBA (green stripe around zero bias) 
is stable and independent on the sign of magnetic field.}
    \label{1n_1}
\end{figure}

Qualitatively the same behavior is demonstrated by another Nb-Co$_3$Sn$_2$S$_2$ junction, demonstrating
a larger interface scattering, see Fig.~\ref{1n_1}(a,b).  The 
blue curve in Fig.~\ref{1n_1}(a) at zero magnetic field and 30~mK demonstrates a well-defined gap 
$\Delta$ = 1.3~meV, denoted by the blue dashed lines, and a subgap structure with a sharp ZBA in the center. 
Again, ZBA is stable even in high magnetic fields, e.g., see the red curve at $B$ = 1.67~T in 
Fig.~\ref{1n_1}(a). The colormap in Fig.~\ref{1n_1}(b) demonstrates stability of ZBA in moderate magnetic 
fields and its independence on the field's sign.

\section{Discussion}

Our main observation is the zero-bias anomaly, which is unusually robust over a wide range of the externally 
applied magnetic field and appears in the proximity-induced charge transport through the junctions involving
the magnetic Weyl semimetal Co$_3$Sn$_2$S$_2$ connecting the superconducting Nb leads. 

In our experiment, superconductivity is proximity induced in an interface vicinity of Co$_3$Sn$_2$S$_2$ WSM, 
which is characterized by the strong spin-orbit coupling and significant intrinsic Zeeman splitting. Note that
the interface possesses an inhomogeneous structure, that could play an important constructive role in forming 
the magnetically stable ZBA. Similar situation was identified in the proximitized semiconductor nanowires 
virtually representing the one-dimensional systems and involving the spin-orbit coupling and the applied 
magnetic field producing the Zeeman splitting. The corresponding tunneling conductance characteristics have 
been progressively studied in such systems both experimentally \cite{mourik2012,das2012,deng2012,chang2015,%
albrecht2016,mzm2,chen2017,gul2017,zhang2017,gul2018,grivnin2019,chen2019} and theoretically 
\cite{lutchyn2010,oreg2010,lee2012,altland2012,plee2012,degottardi2012,degottardi2013,pikulin2012,takei2013,%
loss2013,adagideli2014,chevallier2012,brouwer2012,prada2012,roy2013,defranceschi2014,stanescu2014,%
klinovaja2015,sanjose2016,mzm2,mzm1,trauzettel2018,moore2018,loss2018,vuik2019,lai2019,woods2019,%
stanescu2019,woods2020}. In contrast to the semiconductor nanowire-based experiments, we observe ZBA in the
proximitized magnetic Weyl semimetal, which is a three-dimensional sample in principle, while the 
corresponding chiral edge states reside in its two-dimensional surface. Fortunately, the applicability to 
two-dimensional systems of the main qualitative theoretical statements regarding the Majorana and Andreev 
states in one-dimensional case, is known to be usually justified \cite{flensberg,sato,woods2019,vuik2019}. In 
addition, the problem could acquire some features of the one-dimensional character due to the anisotropy of 
the Fermi-arc surface transport and, possibly, the characteristic structure of surface termination. These 
allow us to use below some of those results for interpreting our experimental findings for ZBA in the 
Nb-Co$_3$Sn$_2$S$_2$ junctions.

A number of mechanisms resulting in ZBA in mesoscopic physics can be ruled out for our experiment. 
This in particular concerns the Oersted field-induced ZBA, when the circular Oersted field, exceeding the 
critical field $H_c$ of a superconductor due to a large current density through an interface of a large area, 
leads to the current driven resistance peaks and ZBA~\cite{zba1,zba2}. Such a possibility can be excluded 
since (i) ZBAs caused by the Oersted field are usually sensitive to an external magnetic field~\cite{zba1}; 
(ii) ZBAs caused by the Oersted field exist almost till the superconductor's $T_c$~\cite{zba1}, in contrast to
the temperature dependence in Fig.~\ref{2n_1}(c). Also,  ZBAs in Fig.~\ref{2n_1}(a) and Fig.~\ref{1n_1}(a) 
cannot arise due to the Kondo effect, since they are symmetric and do not show a double-peak structure under 
the applied magnetic field~\cite{zba1,kondo1,kondo2}. Furthermore, one should point out a possibility for the 
impurity-induced ZBA, which was actively studied with regard to semiconductor nanowires \cite{altland2012,%
degottardi2012,plee2012,pikulin2012,degottardi2013,loss2013,adagideli2014}. A detailed analysis has shown that
disorder-induced ZBA is generally not so stable with varying strength of the Zeeman splitting \cite{loss2013},
as compared to the experimental results in Figs.~\ref{2n_color} and \ref{1n_1}.

At the same time, there are a few possible scenarios resulting in ZBA that yet cannot be excluded from the 
list of relevant physical mechanisms. In general, the topologically nontrivial zero-energy ABSs like Majorana 
bound states, localized at the surface of Co$_3$Sn$_2$S$_2$, could result in the prominent and magnetically
stable ZBA being mainly robust with respect to disorder. An additional reason for ZBA to appear is known to be
associated with the topologically trivial (non-Majorana) near-zero interfaced Andreev modes induced by the 
smooth inhomogeneities \cite{chevallier2012,brouwer2012,prada2012,roy2013,defranceschi2014,stanescu2014,%
klinovaja2015,sanjose2016,mzm2,mzm1,trauzettel2018,moore2018,loss2018,vuik2019,lai2019,woods2019,%
stanescu2019,woods2020}. 

In particular, if there is a spatially inhomogeneous confining potential inside a proximitized nanowire, 
there are subgap ABSs. In the absence of the intrinsic Zeeman splitting as well as the external magnetic 
field, ABSs usually appear as two symmetric subgap conductance peaks. With increasing the 
external magnetic field in the presence of a strong spin-orbit coupling, ABSs can merge to form a single 
zero-energy state producing a single zero-bias conductance peak \cite{chevallier2012,brouwer2012,prada2012,%
roy2013,defranceschi2014,stanescu2014,klinovaja2015,sanjose2016,mzm2,mzm1,trauzettel2018,moore2018,loss2018,%
vuik2019,lai2019,woods2019,stanescu2019,woods2020}. Analogous effects can take place even for highly 
transparent contacts between a superconductor and a chiral channel \cite{sanjose2016}.

Since the chiral edge states are topologically protected in WSM and topologically trivial and 
unprotected ABSs can be, to a certain extent, robust to disorder \cite{stanescu2019}, a similar physical 
mechanism can be assumed to form zero-energy ABSs at the surface of Co$_3$Sn$_2$S$_2$ in proximity with a 
Nb superconductor. 

(i) Although Weyl surface states are two-dimensional, they inherit the chiral property of the Chern insulator
edge states, so a preferable direction is defined by Fermi arcs on a particular crystal surface. 

(ii) The condition of a strong spin-orbit interaction is obviously satisfied for the Co$_3$Sn$_2$S$_2$ 
WSM~\cite{kagome}. A sufficiently large intrinsic Zeeman splitting in the half-metallic Co$_3$Sn$_2$S$_2$
can be tuned by applying the external magnetic field. In the experiment, we observe no evidences of the 
finite-energy Andreev states, meaning that the inherent for the material Zeeman splitting substantially 
exceeds its critical value, above which the finite-energy Andreev states coalesce together and form 
near-zero-energy midgap states.

(iii) An inhomogeneous potential is expected at the surface of Co$_3$Sn$_2$S$_2$, providing a platform for 
confined ABSs. After a mechanical cleavage, one expects to have multiple layer steps in the contact region for
the Nb-Co$_3$Sn$_2$S$_2$ junctions. Cleavage mostly occurs between the Sn and S layers, but the topological 
surface states were revealed at the Sn layers~\cite{kagome_arcs,layers1,layers2}. In addition, it has also 
been shown that there are Co$_3$Sn terraces at the surface of Co$_3$Sn$_2$S$_2$~\cite{terraces}. 

\section{Conclusion}

In conclusion, we have experimentally investigated charge transport through the interface between the Nb 
superconductor and magnetic WSM Co$_3$Sn$_2$S$_2$. Aside from Andreev reflection, we observe several 
additional features, among which the most notable is the prominent subgap ZBA, fairly stable against external 
magnetic fields up to the critical field of Nb. Possible mechanisms for magnetically stable ZBA to appear 
under the conditions studied should include the topological nature of WSM and its interfaces together with the
effects of the strong spin-orbit coupling and large Zeeman splitting, which are intrinsic for the 
half-metallic WSM. As the promising scenario for ZBA observed in the transport measurements, we consider the 
proximity induced zero-energy ABSs interfacing between the superconducting Nb and Co$_3$Sn$_2$S$_2$.

\acknowledgments
We wish to thank V.T. Dolgopolov for fruitful discussions, and S.V~Simonov for X-ray sample characterization.
We gratefully acknowledge financial support partially by the RFBR  (project No.~19-02-00203), RAS, and RF
State task.

\end{document}